\documentclass[%
 aip,
 amsmath,amssymb,
 reprint,%
]{revtex4-1}

\usepackage{graphicx}
\usepackage{dcolumn}
\usepackage{bm}

\usepackage[utf8]{inputenc}
\usepackage[T1]{fontenc}
\usepackage{mathptmx}
\usepackage{textgreek}

\begin{document}

\preprint{AIP/123-QED}

\title[THz intersubband electroluminescence from n-type Ge/SiGe quantum cascade structures]{THz intersubband electroluminescence from n-type Ge/SiGe \\ quantum cascade structures}

\author{David Stark}
\affiliation{Institute for Quantum Electronics,   Department of Physics,  ETH Z\"urich,  Switzerland}
\author{Muhammad Mirza}
\affiliation{James Watt School of Engineering, University of Glasgow, Glasgow G12 8LT, United Kingdom}
\author{Luca Persichetti}
\author{Michele Montanari}
\affiliation{Dipartimento di Scienze, Universita Roma Tre, Roma 00146, Italy}
\author{Sergej Markmann}
\author{Mattias Beck}
\affiliation{Institute for Quantum Electronics,   Department of Physics,  ETH Z\"urich,  Switzerland}
\author{Thomas Grange}
\author{Stefan Birner}
\affiliation{nextnano GmbH, Konrad-Zuse-Platz 8, München 81829, Germany}
\author{Michele Virgilio}
\affiliation{Dipartimento di Fisica “E. Fermi,” Universita di Pisa, Pisa 56127, Italy}
\author{Chiara Ciano}
\affiliation{Dipartimento di Scienze, Universita Roma Tre, Roma 00146, Italy}
\author{Michele Ortolani}
\affiliation{Sapienza University of Rome, Department of Physics, Piazzale Aldo Moro 2, I-00185 Rome, Italy}
\author{Cedric Corley}
\affiliation{IHP - Leibniz-Institut f\"ur innovative Mikroelektronik, Im Technologiepark 25, D-15236 Frankfurt (Oder) Germany}
\author{Giovanni Capellini}
\affiliation{Dipartimento di Scienze, Universita Roma Tre, Roma 00146, Italy}
\affiliation{IHP - Leibniz-Institut f\"ur innovative Mikroelektronik, Im Technologiepark 25, D-15236 Frankfurt (Oder) Germany}
\author{Luciana Di Gaspare}
\author{Monica De Seta} 
\affiliation{Dipartimento di Scienze, Universita Roma Tre, Roma 00146, Italy}

\author{Douglas J. Paul}
\affiliation{James Watt School of Engineering, University of Glasgow, Glasgow G12 8LT, United Kingdom}
\author{Jérôme Faist}
\author{Giacomo Scalari}
\affiliation{Institute for Quantum Electronics, Department of Physics,  ETH Z\"urich, Switzerland}
\email{scalari@phys.ethz.ch}

\begin{abstract}
We report electroluminescence originating from L-valley transitions in \textit{n}-type Ge/Si$_{0.15}$Ge$_{0.85}$ quantum cascade structures centered at 3.4 and 4.9 THz with a line broadening of $\Delta f/f \approx 0.2$. Three strain-compensated heterostructures, grown on a Si substrate by ultrahigh vacuum chemical vapor deposition, have been investigated. The design is based on a single quantum well active region employing a vertical optical transition and the observed spectral features are well described by non-equilibrium Green's function calculations. The presence of two peaks highlights a suboptimal injection in the upper state of the radiative transition. Comparison of the electroluminescence spectra with similar GaAs/AlGaAs structure yields one order of magnitude lower emission efficiency.
\end{abstract}

\maketitle

Since the demonstration of the semiconductor diode laser and the high popularity of Si-based transistor technology, a laser on silicon constitutes a long-standing goal for silicon photonics \cite{liang_recent_2010}. Significant advantages for a Si based laser should result from the high yield manufacturing processes to allow low cost at high volume but also enable low cost photonic systems from photonic integrated circuits. The major problem to realize a Si-based interband laser is the indirect band gap of group IV materials, which leads to a poor radiative recombination rate. Several solutions have been developed to achieve laser action from silicon \cite{boyraz_demonstration_2004,armand_pilon_lasing_2019,wan_tunable_2019,seifried_monolithically_2018,nguyen-van_quantum_2018}. Employing intersubband transitions in quantum cascade structures represent an exciting option because such transitions are independent of the nature of the band gap. 

So far the Quantum Cascade Laser (QCL)\cite{faist_quantum_1994} has only been demonstrated in polar III-V compound semiconductor materials based on transitions between conduction band states. Intersubband electroluminescence, however, due to valence band transitions from non-polar Si/SiGe heterostructures have been observed for mid-infrared \cite{dehlinger_intersubband_2000} and far-infrared wavelengths \cite{lynch_intersubband_2002,paul_progress_2010}. The strained structure produces a complex valence band and the large effective mass ($\sim 0.3m_0$\cite{matmon_sisige_2010}, where $m_0$ is the free electron mass) results in poor gain and therefore no subsequent laser action has been demonstrated. 
Theoretically n-type Ge/SiGe and Ge/GeSiSn material configurations with lower effective mass ($\sim 0.135m_0$) are predicted to be promising candidates to realize a room temperature THz QCL \cite{driscoll_silicon-based_2006,driscoll_design_2007,sun_strain-free_2007,valavanis_material_2011}. Due to the absence of the restrahlen band, lasing transitions above 6 THz should be accessible. Because of the large lattice mismatch between Si and Ge, the growth of such Ge-rich structures on Si wafers is particularly challenging \cite{paul_progress_2010}. Only in the last few years the Ge/SiGe heterostructures reached the quality standard required for this kind of application\cite{grange_atomic-scale_2020}. 

While buried InGaAs/InAlAs QCLs\cite{beck_continuous_2002}, operational in continuous wave and at room-temperature, \textit{de facto} unlocked the mid-IR spectral region in different fields \cite{vitiello_quantum_2015,galli_molecular_2011,klocke_single-shot_2018}, THz QCLs remain limited to operation below room temperatures hindering potential applications. Pulsed Peltier-cooled operation of GaAs/AlGaAs THz QCLs was achieved in 2019 \cite{bosco_thermoelectrically_2019} and recently demonstrated up to 250 K \cite{khalatpour_high-power_2020}. The values of current density and dissipated power in such devices still prevent integration into more complex photonic structures. The high current density, high voltage drop per period and the temperature dependence of the subband lifetimes are intrinsically related to the polar nature of the gain material. Scattering of electrons with LO phonons ultimately limits the population inversion in such devices. 

We recently proposed a Ge/SiGe quantum cascade design where non-equilibrium Green's functions (NEGF) based calculations predict gain up to room temperature \cite{grange_room_2019}. The design is based on a 4-quantum well active region employing a bound-to-continuum transition and involves a total of four subband states. Although such a design approach holds good promise for the demonstration of a laser, it is not the ideal candidate to develop a new THz quantum cascade emitter. The diagonal optical transition leads to broad emission \cite{blaser_far-infrared_2000} and a high voltage drop per period .

\begin{figure*}
    \includegraphics{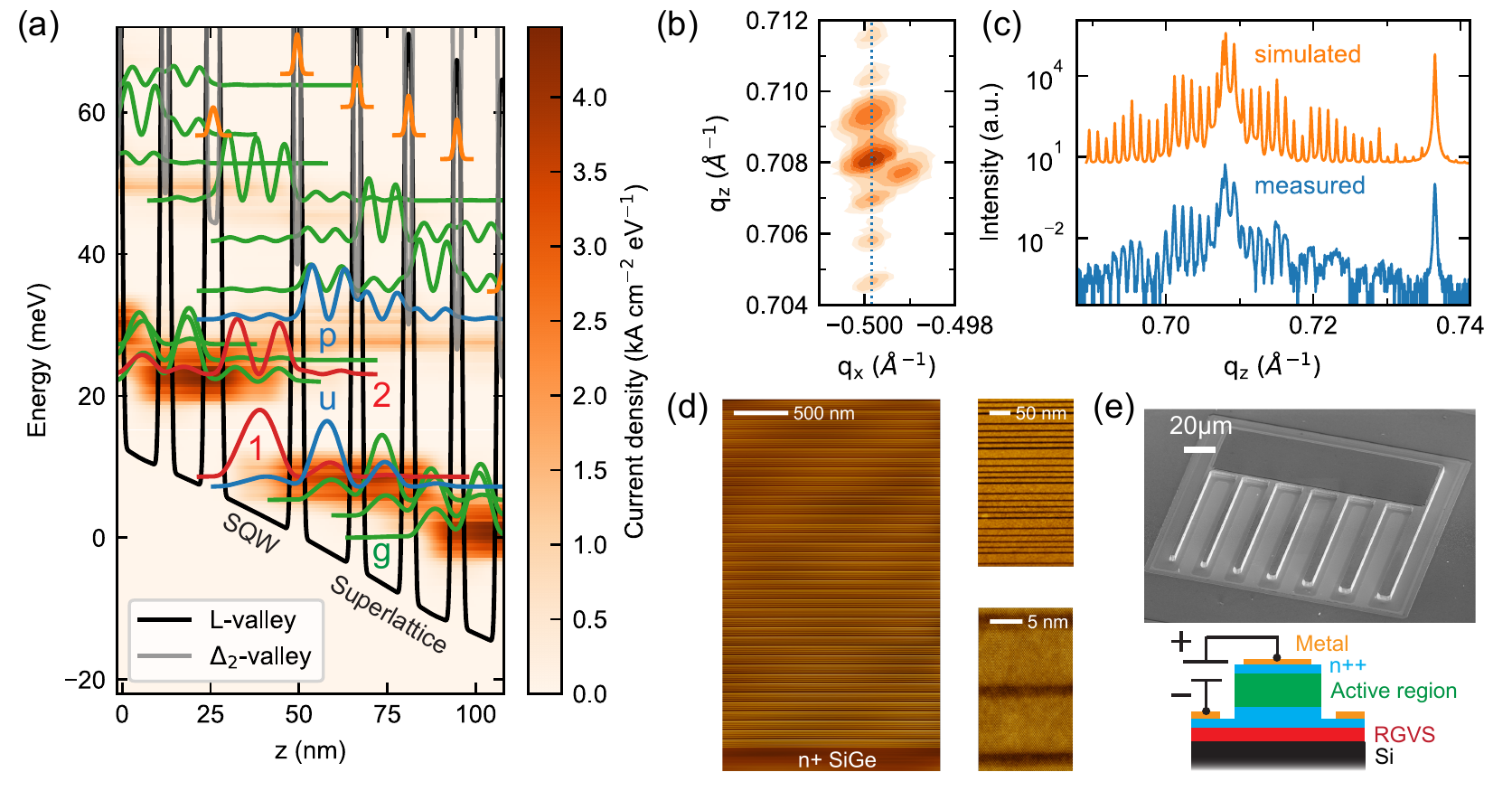}
    \caption{\label{fig:bs_growth}
    (a) Conduction band diagram and Wannier-Stark states for the target SQW design at 22 mV/period computed by NEGF. The solid black and solid grey lines are the potentials in the L- and $\Delta_2$-valleys, respectively. The orange scale shows the position and energy resolved current density at 10 K. The integrated current density is 37 A/$\text{cm}^2$. The nominal period length is 82.4 nm with a sheet doping density of $2.54 \times 10^{10} \text{\,cm}^{-2}$. Starting from the injection barrier the nominal layer sequence with thicknesses in nanometers is \textbf{4.7}/19.9/\textbf{3}/13.9/\textbf{3}/11.6/\textbf{3}/3.8/\underline{3}/3.8/\textbf{3}/9.7. The Ge wells are in standard font, the $\text{Si}_{0.15}\text{Ge}_{0.85}$ barriers are in bold and the phosphorus doped layer is underlined. 
    (b) Reciprocal space map acquired on sample 2306 around asymmetric ($4\bar 2 \bar 2$) Si reflections. (c) Comparison of the measured and simulated (004) XRD rocking curve. The simulated curve is shifted for clarity. (d) STEM image at different magnifications of a SQW heterostructure. The SiGe barriers appear darker. (e) SEM image and schematic diagram of the Ge/SiGe interdigitated diffraction grating. The schematic diagram shows the cross-section of a single grating finger and its bias configuration for the regular growth direction. If the growth direction is reversed the polarity is switched.}
\end{figure*}

To unambiguously demonstrate electroluminescence from a Ge/SiGe quantum cascade structure, we adapted the GaAs/AlGaAs single quantum well (SQW) design reported in Ref \cite{scalari_population_2007}. SQW active regions are not expected to show high optical gain. Instead, the low current density together with the moderate energy drop per period lead to reduced heating of the device. Hence, the unwanted blackbody emission can be reduced. The narrow spectral peak of the vertical intersubband transition should result in a clear signature in the spectrum. In this work, similar GaAs/AlGaAs structures with the same expected emission energy \cite{rochat_far-infrared_1998} are used for a quantitative benchmark comparison with the Ge/SiGe results.

The conduction band profile in the L-valley and energy resolved current density of the target design is shown in Fig.~\ref{fig:bs_growth}(a). The main vertical transition between states $|2\rangle$ and $|1\rangle$ is expected at an energy of 14.5 meV at 22 mV/period with a current density of 37 A/$\text{cm}^2$. The NEGF calculations are performed with the \textit{nextnano.NEGF} simulation package \cite{grange_electron_2014,grange_contrasting_2015}. The effective mass along the confinement direction of $m^{*} = 0.135 m_0$ is used \cite{gupta_room_2018,ciano_control_2019}. In this strain-compensated design we employ tensile-strained $\text{Si}_{0.15}\text{Ge}_{0.85}$ barriers. Such a Si-content leads to a smaller confinement offset in the L-valley $\Delta E_\text{L} = 80 \; \text{meV}$ compared to the $\text{GaAs}/\text{Al}_{0.15}\text{Ga}_{0.85}\text{As}$ material system with $\Delta E_\Gamma = 125 \; \text{meV}$ \cite{yi_probing_2010}. This low Si content separates energetically the parasitic $\Delta_2$-valley states from the L-valley states to avoid potential leakage or charge trapping by the barriers. Note that the simulation model accounts for interdiffused Ge/SiGe interfaces. The interface properties are based on atom probe tomography measurements \cite{grange_atomic-scale_2020}. Here the interface roughness (root mean square) $\Delta = 0.11$ nm and the in-plane correlation length $\Lambda_{||} = 7.0$ nm are used.

\begin{table}[b]
    \caption{\label{tab:sample_overview} 
    Overview of the quantum cascade structures: Sample ID, nominal sheet doping density $n_\text{2D}$, period length $L_\text{p}$ measured by XRD and the number of periods $N_\text{p}$.}
    \begin{ruledtabular}
    \begin{tabular}{cccc}
    Sample ID & $n_\text{2D} \big(10^{10}\text{cm}^{-2}\big)$ & $L_\text{p} \big(\text{nm}\big)$ & $N_\text{p}$\\
    \hline
    2306 & 5 & 85.4 & 51\\
    2307 & 7 & 83.3 & 51\\
    2315\footnote{reverse growth direction} & 7 & 87.8 & 51\\
    \hline
    S1353\footnote{reference GaAs/$\text{Al}_{0.15}\text{Ga}_{0.85}\text{As}$ sample}
    & 2.9 & 97.7 & 35
    \end{tabular}
    \end{ruledtabular}
\end{table}

The strain-compensated structures were grown on $\text{Si}_{1-x}\text{Ge}_{x}$ reverse graded virtual substrate (RGVS) on Ge/Si(001) substrates by ultrahigh vacuum chemical vapour deposition (UHV-CVD) \cite{seta_narrow_2012,grange_atomic-scale_2020,persichetti_intersubband_2020}. The RGVS is 3.7 $\mu\text{m}$ thick where the last 1.3 $\mu\text{m}$ is a constant composition $\text{Si}_{0.03}\text{Ge}_{0.97}$ buffer. The quantum cascade periods are embedded between a 400 nm bottom- and a 25 nm top-contact. Both contacts consist of phosphorous doped $\text{Si}_\text{0.03}\text{Ge}_{0.97}$ with $N_\text{D} \approx 2 \times 10^{19}\; \text{cm}^{-3}$. The active region consists of 51 periods of the SQW region design (see Fig.~\ref{fig:bs_growth}(a)) for a total thickness of $4.2 \;\mu$m.
The main sample parameters including the GaAs/AlGaAs reference discussed in this work are listed in Table~\ref{tab:sample_overview}. The ($4\bar 2 \bar 2$) x-ray diffraction (XRD) reciprocal space map of sample 2306 depicted in Fig.~\ref{fig:bs_growth}(b) shows the lattice matching of the active region to the $\text{Si}_\text{0.03}\text{Ge}_{0.97}$ buffer as evidenced by the alignment of the superlattice peaks with the $\text{Si}_\text{0.03}\text{Ge}_{0.97}$ signal along the vertical dotted line. 
The active region period length $L_\text{p}$ and its reproducibility is derived from 1D XRD rocking curves, see Fig.~\ref{fig:bs_growth}(c). The scanning transmission electron microscopy (STEM) in Fig.~\ref{fig:bs_growth}(d) shows the quality of the UHV-CVD growth. The threading dislocation density is estimated to be on the order of
$10^{6}\;\text{cm}^{-2}$, a state-of-the-art value for this material system \cite{skibitzki_reduction_2020}. As the threading dislocations act as acceptor-like states, the phosphorous density in the Ge/SiGe structures is larger than the free carrier concentration targeted at the design stage in Fig.~\ref{fig:bs_growth}(a) \cite{claeys_electrical_2009}. Moreover, it is not well known how many carriers per unit length are trapped by a threading dislocation, thus the active carrier concentration is affected by some uncertainty.

The samples were processed into deeply etched diffraction gratings with linearly chirped periods (23 - 28 $\mu$m). The top and bottom metallizations are interdigitated to uniformly pump the device area. To reduce the amount of defects per device, the area of the etched mesa is kept small (5.5\% of the focal spot of the detector which is $450\times450 \;\mu\text{m}^2$). A SEM image of the processed device and the schematic diagram of the cross-section of the device are illustrated in Fig.~\ref{fig:bs_growth}(e). The silicon based samples were etched using inductively coupled plasma (ICP) etching with SF$_\text{6}$/C$_\text{4}$F$_\text{8}$ chemistry \cite{mirza_nanofabrication_2012}. Ohmic contacts were formed on the highly doped semiconductor material using deposited Ni metal annealed at 330 $^\circ$C for 30 s to form NiGe \cite{gallacher_ohmic_2012}. The linearity of the contact resistances were evaluated with circular transmission line test structures at 10 K (see Supplementary material). The GaAs/AlGaAs reference structure was processed into gratings with identical dimensions. It was etched using ICP with SiCl$_\text{4}$/N$_\text{2}$ chemistry and metallized with Ge/Au/Ni/Au serving as well as a self aligned etch mask. 

Electroluminescence measurements were performed in a home-made, vacuum Fourier transform infrared (FTIR) spectrometer equipped with a He-cooled Si bolometer (IR Labs). The samples were mounted in a He flow cryostat. The heat sink temperature was kept at 5 K. The setup was pre-aligned and phase calibrated using a vertically emitting single mode THz QCL
\cite{bosco_patch-array_2016} driven below threshold. 
Current was injected following a micro-macro pulse scheme to minimize heating effects. Generally, 627 micro-pulses of 960 ns length (equivalent to a duty-cyle of $D=50\%$) with macro-pulse frequency of 415 Hz are used. The interferograms were acquired in step-scan mode with a lock-in amplifier with 2 s integration time per point and a spectral resolution of $\Delta E = 0.82$ meV (198 GHz). Multiple interferograms with the same experimental conditions were averaged yielding typically 8-24 hours of total integration time. The stabilities of the in-phase component, quadrature component and the heat sink temperature were verified continuously during the acquisition time.

The voltage-current (VI) and electroluminescence intensity-current (LI) characteristics measured at 5 K for the Ge/SiGe samples and the GaAs/AlGaAs reference sample are visible in Fig.~\ref{fig:LIV} (VI curves as a function of the temperature up to 290 K are reported in the Supplementary Material). The observed Ge/SiGe VI curves are compared to the NEGF simulation for  sample 2307; the measured current density exceeds the simulated one by a factor of 5, that is consistent with theoretical lifetimes predictions, as discussed futher below.  Comparing the VI curves of the Ge/SiGe samples to the GaAs reference measurement, the current density  for the same applied electric field is one order of magnitude larger. Additionally, the transport of the Ge/SiGe samples does not show signs of negative differential resistance (NDR) as clearly observed in the reference sample at a current density of 42 $\text{A}/\text{cm}^2$.
From Fig.~\ref{fig:LIV}(a), it can be seen that the spectra of the Ge/SiGe samples are acquired at low injection currents (< 200 mA or 1.8 kA/$\text{cm}^2$) to prevent any blackbody emission overwhelming the intersubband signal.

\begin{figure}[t]
    \centering
    \includegraphics{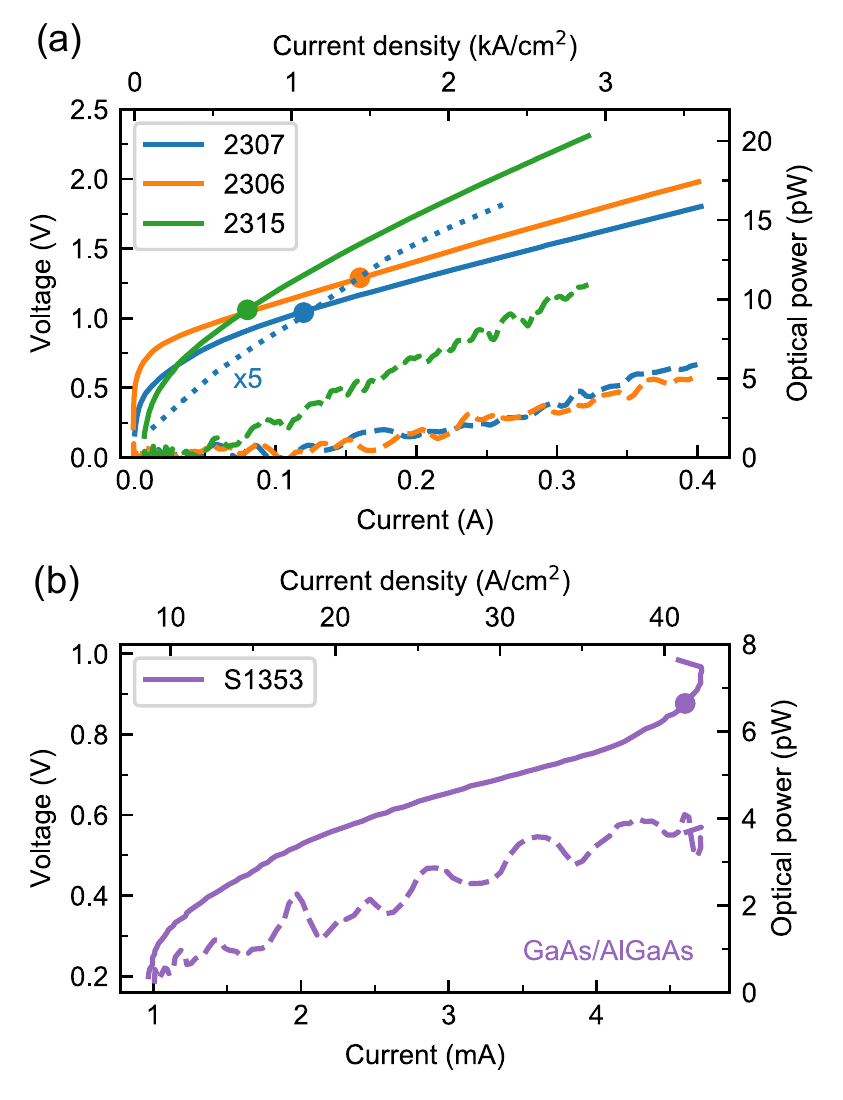}
	\caption{Light-current-voltage (LIV) characteristics at $T = 5\; \text{K}$ for the Ge/SiGe samples (a) and the GaAs/AlGaAs reference sample (b). The solid and dashed lines are the VI- and LI-curves respectively. The dots indicate the operation points where electroluminescence spectra are acquired. The dotted line in (a) is the VI curve computed by NEGF for sample 2307, where the current is multiplied by 5. Note that for the reference measurement a micro-pulse duty-cycle of $D=95\%$ was used.}
	\label{fig:LIV}
\end{figure}

\begin{figure*}[t]
    \centering
    \includegraphics[width=1\linewidth]{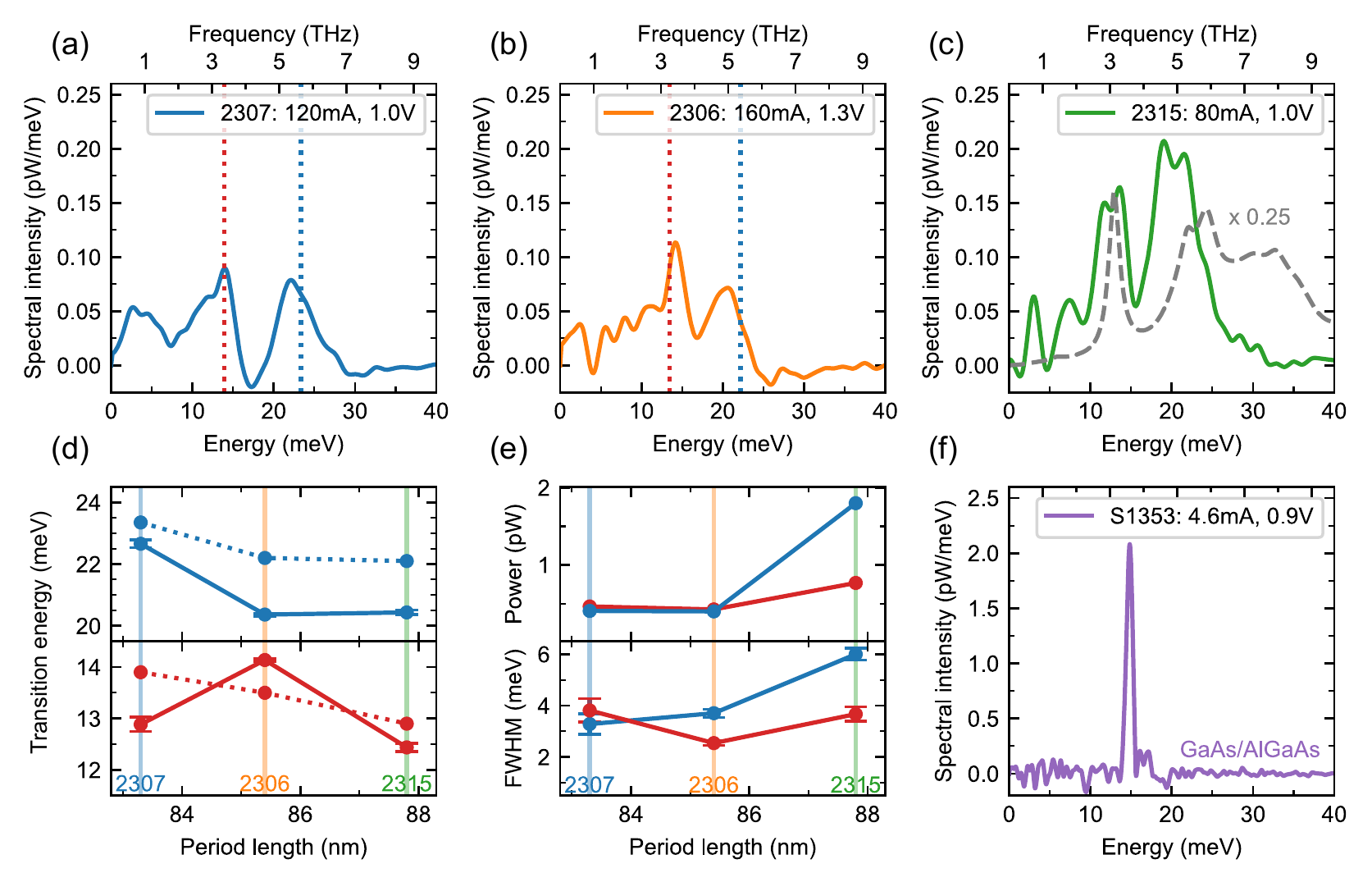}
	\caption{Electroluminescence spectra of the three Ge/SiGe samples (a), (b) and (c). The dotted lines correspond to the simulated peak positions. (c) The gray dashed line shows the theoretically calculated electroluminescence spectrum. (d) The simulated peak positions are compared with the measured peak positions. (e) Measured FWHMs and integrated optical peak power as function the period length. (f) GaAs/AlGaAs reference spectrum with a transition energy of 14.8 meV, FWHM of 0.7 meV and integrated peak power of 2.5 pW. Note that this measurement is acquired with spectral resolution $\Delta E = 0.24$ meV and micro-pulse duty-cycle $D=95\%$.}
	\label{fig:EL_spectra}
\end{figure*}

Typical spectra measured on the three Ge/SiGe samples are shown in Fig.~\ref{fig:EL_spectra}(a)-(c). Two peaks centered around 13 meV and 22 meV appear distinctly from the spectra. We identify the first peak as the vertical transition between states $|2\rangle$ and $|1\rangle$, see Fig.~\ref{fig:bs_growth}(a). We attribute the second peak to a transition from a higher lying energy state $|\text{p}\rangle$ to the upper state of the miniband $|\text{u}\rangle$, which extracts the electrons from the SQW. In Fig.~\ref{fig:EL_spectra}(c) we report as well the predicted electroluminescence signal using NEGF. Two peaks are predicted, in fair agreement with the measurements: the presence of the second peak highlights the role of hot electrons due to the absence of strong inelastic scattering (LO phonons). The measured spectra are fitted to a double Lorentzian function and the extracted transition energies are compared to the simulated peak positions as a function of the period length, see Fig.~\ref{fig:EL_spectra}(d). The measurement and simulations agree to within 2 meV. Note that in simulations we account for the different period lengths obtained by XRD and the nominal doping density is used (see Table~\ref{tab:sample_overview}). The small discrepancy of the peak positions may be due to some uncertainty related to the material parameters of the Ge/SiGe system (band-offset and electron effective mass), as well as to the exact knowledge of the actual doping profile. 
In Fig.~\ref{fig:EL_spectra}(e) the extracted FWHM and the integrated peak power of the individual transitions are shown. The observed FWHMs agree with optical absorption measurements reported in Ref \cite{virgilio_physical_2014,ciano_control_2019}.
Considering sample 2306 the line broadening for both peaks is $\Delta f/f \approx 0.2$. The FWHM of the $|2\rangle$ to $|1\rangle$ transition is 2.5 meV, which is 1 meV larger than what is predicted by NEGF. In the context of the Kazarinov-Suris density matrix model for transport in quantum cascade structures \cite{RN1300,SirtoriIEEEresTunneling1998}, this linewidth corresponds to a dephasing lifetime $T_2\simeq 0.3$ ps.
It has to be mentioned that a control sample with the doping placed homogeneously over four quantum wells of the injector and processed in identical grating has given, as expected, a low intensity, broad signal with no significant spectral features. Such measurements and additional reverse bias measurements are reported in the Supplementary Material.

The spectrum of the GaAs/AlGaAs reference sample is reported in Fig.~\ref{fig:EL_spectra}(f). It is acquired at the maximum current value before the NDR. The peak is centered at 14.8 meV with a FWHM of 0.7 meV ($\Delta f/f \approx 0.05)$ and integrated peak power of 2.5 pW. 
Only one peak appears due to proper electronic injection into the upper state $|2\rangle$  and an accurate doping level. However, above the NDR two peaks have been reported due to broad injection into high-energy states \cite{rochat_far-infrared_1998}, where the corresponding spectrum resembles the spectra of the Ge/SiGe samples. The smaller confinement offset in our $\text{Si}_{0.15}\text{Ge}_{0.85}$ design energetically lower the parasitic state $|\text{p}\rangle$. The weaker electron-phonon interaction in non-polar Ge/SiGe active region leads to a larger electronic excess temperature \cite{grange_room_2019,bagolini_disentangling_2020}. Thus, the transport through higher lying energy states is non negligible and leads to a second optically active transition. Additionally, the doping level in the active region and the contacts of our structures are not optimized. This could cause a non-homogeneous electric field across the 51 periods of the cascade and ultimately impair the electronic injection.

In the following the non-radiative lifetimes of the upper states $|2\rangle$ and $|p\rangle$ are estimated, considering the spectrum of sample 2315 and using the GaAs reference to extract the collection efficiency. The electroluminescence power can be written as
\begin{eqnarray}
    P_\text{opt} = 
    \eta_\text{inj} \eta_\text{coll}  \eta_\text{rad}
    \frac{h\nu}{e} N_\text{p} D\cdot I,
\end{eqnarray}
where $\eta_\text{coll}$ is the collection efficiency for the setup and the grating, $\eta_\text{rad} \approx \tau_\text{nr}/\tau_\text{sp}$ the radiative efficiency, $h\nu$ the photon energy, $D$ the duty-cycle, $I$ the current and $e$ the absolute charge of an electron. In the case of GaAs, we assume unitary current injection $\eta_\text{inj} = 1 $ (NDR well visible), we use  $\tau_\text{nr} = 12.2\;\text{ps}$\cite{scalari_population_2007} and, with the computed spontanoeus emission lifetime $\tau_\text{sp} = 7.4 \;\mu\text{s}$  we can extract a collection efficiency of $\eta_\text{coll} \approx 7\times 10^{-4}$. 
For the Ge/SiGe structure we assume that the total power is given by sum of the two peak powers (see supplementary material). The injection efficiencies into states $|2\rangle$ and $|p\rangle$ computed with NEGF simulations are $\eta_\text{inj,2} = 0.38$ and $\eta_\text{inj,p} = 0.37$. Using the computed spontaneous emission lifetimes $\tau_\text{sp,2} = 15.9 \;\mu$s and $\tau_\text{sp,p} = 7.1 \;\mu$s, we obtain $\tau_\text{nr,2} = 1.9$ ps and $\tau_\text{nr,p} = 1.3$ ps. Consequently, the radiative efficiency is $\eta_\text{rad} \approx 10^{-7}$. The lifetimes predicted by NEGF modelling are $\tau_\text{nr,2}^\text{sim} = 12$ ps and $\tau_\text{nr,p}^\text{sim} = 6$ ps, roughly a factor of 5 higher. We stress that the deduced values for the lifetimes constitute a lower limit; the actual injection efficiency is not known due to the presence of threading dislocations and, as discussed above, to possible inhomogeneities in the electric field across the whole sample length. 
Nevertheless, a lower value of $\tau_\text{nr,2}$ with respect to the NEGF simulation may also result from the larger effective doping density adopted and the higher electronic temperature. In fact, NEGF simulations suggest that Coulomb scattering prevails over interface roughness in the determination of the lifetimes and of the linewidth.
%
%

In conclusion, THz intersubband electroluminescence from n-type Ge/$\text{Si}_{0.15}\text{Ge}_{0.85}$ quantum cascade structures has been demonstrated from three different epitaxial layers. The spectral features agree well with theoretical predictions based on the non-equilibrium Green's function simulations. The emitters have been benchmarked against a similar GaAs/AlGaAs structure with identical device geometry. The Ge/SiGe emission efficiency is one order of magnitude lower. This is attributed to a suboptimal injection of the electrons into the upper state of the radiative transitions and to a lower upper state lifetime.
Future improvements will come from a more accurate control of the doping profile, a better understanding of its interaction with the threading dislocations and from a further reduction of their density. Finally it would be useful to clarify whether the $\Delta_2$-states confined in the barriers influence the electronics dynamics.  
To achieve laser action, we will leverage on the progress made in the material growth and structure modelling by employing higher gain structures based on 4-quantum wells \cite{grange_room_2019} featuring a diagonal optical transition and embedded in double metal waveguides \cite{gallacher_design_2020}.
\\\\
This project was funded by the European Union’s Horizon 2020 research and innovation programme under Grant Agreement No. 766719 (FLASH) and European Research Council Consolidator Grant (724344) (CHIC). The data that support the findings of this study are available from the corresponding author upon reasonable request. The authors would like to thank Andres Forrer, Bo Meng and Hans Sigg for useful discussions and support.
\bibliography{BIB_david}

\end{document}